# Pressure-induced electronic topological transition and superconductivity in topological insulator $Bi_2Te_{2.1}Se_{0.9}$

*Lei Kang,*[1] *Zi-Yu. Cao*[2*]*, and Bo Wang*[3]

[1]*College of Medical Information Engineering, Guangdong Pharmaceutical University, Guangzhou, 510006, China*

[2] *Center for Quantum Materials and Superconductivity (CQMS) and Department of Physics, Sungkyunkwan University, Suwon 16419, Republic of Korea*

[3] *School of Applied Physics and Materials, Wuyi University, Jiangmen 529020, China*

[*] Correspondence should be addressed to ZYC (zy.cao317@gmail.com)

**One approach to discovering topological superconductor is establishing superconductivity based on well-identified topological insulators. However, the coexistence of superconductivity and topological state is always arcane. In this paper, we report how pressure tunes the crystal structure, electronic structure, and superconductivity in topological insulator $Bi_2Te_{2.1}Se_{0.9}$. The first pressure-induced structure transition start from 8.4 GPa, followed by the structure sequence of *R*-3m - *C*2/m - *C*2/c - *Im*-3*m* up to 37 GPa. Superconductivity starts to present at 2.4 GPa with the $T_c$ around 6.6 K. Moreover, at around 2.5 GPa, the abnormal changes of *c*/*a* and the full width at half maximum (FWHM) of mode indicate the occurrence of electronic topological transition (ETT). These results suggest that ETT highly promotes the superconductivity in topological insulator $Bi_2Te_{2.1}Se_{0.9}$ in the low-pressure rhombohedral phase. Our work clarifies the complex electronic structure in $Bi_2Te_{2.1}Se_{0.9}$ and sheds light on the new candidate for superconductivity in topological insulators.**

Topological insulators are new states of quantum matter that have time-reversal symmetry protected surface or edge states residing in the bulk insulating gap and are, therefore, robust with non-magnetic impurities and external environment.[1-3] By chemical doping or applying pressure, the superconductivity can be accidentally produced in the topological insulators with primordial crystallographic structure. And it provides an alternative way to create a coexistence region of bulk superconductivity and topological state in a single crystal, which may offer a path to discovering the topological superconductor.[4,5] $Bi_2Te_3$ and its isostructural compounds $Bi_2Se_3$ and $Sb_2Te_3$ are prototypical precursors.[6-8] The superconductivity was successfully achieved in copper intercalated $Bi_2Se_3$ at ambient pressure, resulting in intense discussions in the community on the topological superconducting nature of $Cu_xBi_2Se_3$.[4] From the point-contact spectra, a zero-bias conductance peak, which is deemed to be the attribution of Majorana fermions, was discovered in $Cu_xBi_2Se_3$.[9] This observation is supported by theory, [10,11] but, argued by other point-contact spectra measurement, suggesting the zero-bias conductance peak can be tuned by the contact barrier strength.[12] ARPES measurements revealed the existence of nontrivial surface states.[13] Meanwhile, in the dc magnetization measurements, their observation is consistent with the spin-triplet pairing superconductivity with odd parity.[14] However, The STM and specific heat study suggest the electron pairing in $Cu_xBi_2Se_3$ possesses a fully gapped s-wave symmetry, departure with the odd-parity superconductor proposed by the theory.[15-17] Applying external pressure without introducing impurities can also induce superconductivity into the phase protected by the topological surface state.[18,19] The superconductivity can be reviled in topological insulators $Bi_2Te_3$ and $Sb_2Te_3$ without structural phase transition at the pressure of the topologically nontrivial band structure exist.[18,20-22] However, performing further experiments to investigate the superconducting mechanism under pressure is quite difficult. The superconductivity nature of these compounds is still merging further study.

$Bi_2Se_3$ and its isostructural compounds $Bi_2Te_3$ and $Sb_2Te_3$ are well-known thermoelectric materials.[23] Meanwhile, they show rich physical properties under pressure, such as electronic topological transition (ETT). ETT or Lifshitz transition, proposed by Lifshitz, can be driven by an external parameter such as pressure, temperature, doping, or magnetic field, resulting in variations in the topology of the Fermi surface without symmetry breaking.[24] It is characterized as a 2.5 order transition that the thermodynamic potential and its first derivatives are not affected by an ETT, but the second or third derivatives are expected to vary in the vicinity of the ETT. Recent high-pressure powder XRD measurements have proved a pressure-induced ETT in $Bi_2Te_3$ and $Sb_2Te_3$ at around 3.2 GPa and 3.5 GPa.[25-27] It is worth noting that the pressure-induced superconductivity in $Bi_2Te_3$ and $Sb_2Te_3$ are present at the pressure of 3.2 (1.0) and 4.0 (1.2) GPa, respectively,[18,20,28] which is comparable with the ETT pressure within the uncertainty of the pressure step and measurements. The answer to whether the ETT and superconductivity are coupled may shed light on some urgent questions related to superconductivity and topological superconductivity in these compounds.

In this paper, we tune the superconductivity by selenium substitution for tellurium in $Bi_2Te_3$, then track down how ETT varies with pressure. $Bi_2Te_{2.1}Se_{0.9}$ is the solid solution of $Bi_2Te_3$ and $Bi_2Se_3$, and they share a similar structure and topological state.[29] Combined with the X-ray measurement, the sharp drop of resistance to zero at 2.4 GPa indicates that the superconductivity of $Bi_2Te_{2.1}Se_{0.9}$ is not caused by structure transition. We surprisingly found that the pressure-induced ETT is pinned with the superconductivity. Moreover, the structural evolution and vibrational modes properties of $Bi_2Te_{2.1}Se_{0.9}$ under pressure are also investigated and discussed.

Figure 1 shows the temperature-dependent resistance of $Bi_2Te_{2.1}Se_{0.9}$ at various pressures up to about 34.5 GPa. The resistance taken at ambient pressure is shown in the inset. As a narrow-band-gap semiconductor, $Bi_2Te_{2.1}Se_{0.9}$ exhibits metallic behaviors at ambient pressure, which is similar to its mother compound $Bi_2Te_3$.[30] When pressure is increased to about 2.4 GPa, a clear drop of resistance to zero occurs with a narrow transition width of about 0.5 K, which indicates that a superconducting transition emerges in $Bi_2Te_{2.1}Se_{0.9}$ with the onset of transition temperature $T_c$ of 6.6 K. Here, the $T_c$ is defined as the intersection of two extrapolated lines shown in the Figure. With further increasing pressure to 18.3 GPa, a high $T_c$ phase gradually emerges with a $T_c$ of 8.9 K. Under the pressure of 22.9 GPa, only a high $T_c$ phase with sharp transition can be observed. The emergence of the high $T_c$ phase is induced by structural transition, which will be discussed later. It seems that the two superconducting phases coexist in the pressure range of about 18.3-22.9 GPa. The coexistence of superconducting phases can also be found in other materials.[18] With further compression, the $T_c$ is insensitive to pressure up to 34.5 GPa, which is the highest pressure in our electrical transport measurement. It is worth noting that at 28.9 GPa, the normal state resistance curve in the low-temperature region shows an upturning behavior. A hump-like feature near $T_c$ is often observed in disordered superconductors, which can be ascribed to the development of SC puddles surrounded by a normal state.[31,32]

In order to explore the structural behavior and the relationship between crystal structure and superconductivity of $Bi_2Te_{2.1}Se_{0.9}$ at high pressures, we performed high-pressure Raman scattering and XRD studies on $Bi_2Te_{2.1}Se_{0.9}$ up to 27.0 and 36.3 GPa, respectively. Raman spectroscopy is a powerful tool to study the vibrational properties and electron-phonon coupling, which has been widely used in $Bi_2Se_3$ and $Bi_2Te_3$.[33] For $Bi_2Se_3$, $Bi_2Te_3$, $Sb_2Te_3$, and their solid solutions, the space group at ambient pressure belongs tp *R*-3m. There are three quintuple layers, stacked by van der Waal's forces, in a unit cell; each quintuple layer is an alternate arrangement of five atomic layers. Te(1)-Bi-Te(2)-Bi-Te(1), and Te atoms exhibit two different chemical environment (Te(1), Te(2)). The chemical bonding between Bi and Te(2) is covalent, while it exhibits slightly ionic between Bi and Te(1).[34,35] For the $Bi_2Te_{3-x}Se_x$ compounds, Se atoms, showing more electronegative than Te atoms, will occupy the Te(2) sites and then display random replacements of atoms at Te(1) sites.[34,35] For $Bi_2Te_{2.1}Se_{0.9}$, the doping amount of Se is small so Se atoms will occupy Te(2) sites only.

Figure 2 depicts the Raman spectra of $Bi_2Te_{2.1}Se_{0.9}$ at different pressures. Based on the structural motif, 15 dynamical modes exist at the center of the Brillouin zone, among which 12 are the optical branch and the other 3 are the acoustic branch. However, due to Raman inactive or insignificant scattering cross section, only three Raman active modes can be observed, which are assigned as $A_{1g}^{1}$(62 cm$^{-1}$), $E_{g}^{2}$ (104 cm$^{-1}$), and $A_{1g}^{2}$ (141 cm$^{-1}$).[36] The schematic diagram of these vibrational modes is shown in Figure 2(b). $E_g$ modes correspond to atom vibrations in the plane of the layers, while the $A_g$ modes correspond to vibrations perpendicular to the layers.[37] When pressure is increased to 11.8 GPa, abrupt changes occur in the Raman pattern. For example, the $A_{1g}^{1}$ and $A_{1g}^{2}$ modes disappear, and the intensity of $E_{g}^{2}$ shows significant reduction, which indicates a phase transition. Figure 2(b) shows the pressure dependence of vibrational frequencies of the three Raman modes. The Raman modes shift to high frequency with increasing pressure due to the decrease in bond distances and increase in effective force constants.[38] With further compression to 20.7 GPa, no Raman modes can be observed, suggesting the occurrence of a high symmetrized structure. Interestingly, the FWHM of $A_{1g}^{1}$ with pressure exhibits distinct features showing a maximum at around 3.0 GPa in the initial structure. It is easily reminiscent of the recent report in $Bi_2Te_3$ and $Sb_2Se_3$ that the local maximum of the FWHM in $A_{1g}^{1}$ phonon is an indicator of ETT.[39,40]

The high-pressure XRD patterns of $Bi_2Te_{2.1}Se_{0.9}$ up to 36.3 GPa are depicted in Figure S1. X-ray diffraction is believed as a direct and efficient tool to probe the changes in the long-range structure of the crystal. With increasing pressure, the diffraction peaks shift to higher two-theta angles. This can be explained by the reduced distances of crystal planes and the shrinkage of unit cell volume with applied pressure. The structure motif seems robust with pressure below 8.4 GPa. Once the pressure is above 8.4 GPa, there are several changes in the XRD patterns, including the number, intensity, and shape of the diffraction peaks, indicating a structural transition. Upon further compression to 13.3 GPa, the original strongest peak (marked by a down-facing arrow) of phase I disappears, and the new peak of phase II (marked by an up-facing arrow) becomes the strongest peak, indicating that the transformation is complete and the existence of pure phase of phase II. When pressure is increased up to 13.8 GPa, a peak marked by a solid circle emerges, which indicates another phase transition and the emergence of phase III. At 17.7 GPa, a new peak marked by a rhombus appears, indicating the emergence of phase IV. With pressure increased to 22.9 GPa, the diffraction peaks of phase II disappear completely. However, the coexistence of phases III and IV persists to the highest pressure, 36.3 GPa, in our experiment. Upon releasing pressure, the crystal structure returns to phase I, suggesting the reversibility of the phase transitions. The structural evolution of $Bi_2Te_{2.1}Se_{0.9}$ is consistent with those of $Bi_2Te_3$ and $Bi_2Se_3$,[20,30,41,42] with slightly different high-pressure mixed phases. Both Raman and XRD measurements provide consistent evidence for the pressure-induced phase transitions.

We refined the XRD patterns to obtain the accurate crystal structures of various $Bi_2Te_{2.1}Se_{0.9}$ phases. Figures 3(a)-(d) show the Rietveld refinements of XRD patterns at different pressures, and the crystal structures of the four phases are shown in the inset. Figure 3(a) shows the refinement at 0.6 GPa, and it can be indexed with the space group *R*-3m, which agrees well with the reported result.[36] At 13.3 GPa, the crystal structure of phase I transforms into that of II completely. The refinement [Figure 3(b)] of the XRD pattern at 13.3 GPa resolves a pure phase with *C*2m symmetry. When pressure is increased to 21.2 GPa, according to the XRD pattern, we can see XRD peaks from phases

II, III, and IV, indicating that the material is a mixture of three phases (II+III+IV). The refinement [Figure. 3(c)] at 21.2 GPa shows the mix of the three high-pressure phases, which agrees with the observed XRD pattern, and phase III has a *C*2/c symmetry. With further compression to 36.3 GPa, a small peak at around $2\theta$=12.0°, which belongs to phase III, is observed. We refine the pattern at 36.3 GPa by excluding this peak and the refinement [Figure. 3(d)] shows that phase IV can be indexed to a bcc unit cell (occupancy: 0.4 for Bi and 0.6 for Te and Se) within this structure. Bi, Te, and Se atoms are disordered to randomly share the bcc lattice sites, forming a Bi-Te-Se substitutional alloy. A complete pressure-temperature phase diagram is well established, shown in Figure 5.

Figure 4(a) shows the evolution of lattice parameters a and c in $Bi_2Te_{2.1}Se_{0.9}$ as a function of pressure up to about 13.0 GPa. Figure 4(c) shows the pressure dependence of the unit cell per atom up to 36.3 GPa. We can observe obvious discontinuity at several pressure points where phase transitions occur. Recent high-pressure studies reveal the pressure-induced electronic topological transition in $Bi_2Te_3$, $Bi_2Se_3$, and $Sb_2Te_3$.[25-27,39] ETT is characterized as a 2.5 order transition,[24] which is driven by pressure, temperature, doping, etc., resulting in variations in the topology of the Fermi surface that changes the electronic density of the surface near the Fermi energy. No discontinuity of the volume(first derivative of the Gibbs free energy ) can be seen. Still, variations of the second derivative of the Gibbs free energy, such as compressibility, are expected. Moreover, phonon softening and transport properties anomalies may also be observed during ETT.[43] It is shown that there is a clear change in the *c*/*a* ratio, but without volume discontinuity around the ETT in the $Bi_2Te_3$, $Bi_2Se_3$, and $Sb_2Te_3$ under pressure.[25-27,39] We also observed similar changes in $Bi_2Te_{2.1}Se_{0.9}$ at about 2.5 GPa [Figures 4(b) and 4(c)]. To confirm whether an ETT occurs in $Bi_2Te_{2.1}Se_{0.9}$, we performed a linearization of the BM-EOS vs. the Eulerian strain:

$$H = B_0 + \frac{3}{2}B_0(B_0^{'} - 4)f_E \quad (1)$$

Where $B_0$ denotes the isothermal bulk modulus, $B_0^{'}$ is the first pressure derivative of the bulk modulus,

$$H = \frac{P}{3f_E(1+2f_E)^{5/2}} \tag{2}$$

is the reduced pressure, and

$$f_E = \frac{X^2 - 1}{3} \tag{3}$$

is the Eulerian strain, with

$$X = \left(\frac{V_0}{V}\right)^{1/3} \tag{4}$$

where $V_0$ is the volume per atom at ambient pressure, and $V$ is the volume per atom at pressure P given in GPa.

Without the ETT, the reduced pressure H vs. the Eulerian strain $f_E$ should be linear. Figure 4(d) shows the reduced pressure H as a function of the Eulerian strain $f_E$. We can see a clear change at about 2.5 GPa, indicating the presence of an ETT. Figures 4(e) and 4(f) show the H($f_E$) plot for the a($H_a$) and c($H_c$) axes, respectively. In this case, we replace Eq.(4) either by $X_a = (a_0/a)$ or $X_c = (c_0/c)$. The kink present in the slop of $H_a$, however, is absent in the pressure dependence of $H_c$, providing an anisotropy nature in the ETT. It indicates that the compressibility perpendicular to the layers shows great change after ETT, while the compressibility parallel to the layers is not. The resulting is self-consistent with the abnormal changes of $c/a$ shown in figure 4 (b). The changes in the slopes of H and $H_a$ vs. pressure give solid evidence for the pressure-induced ETT at about 2.5 GPa in $Bi_2Te_{2.1}Se_{0.9}$, which is consistent with the changes in the slope of FWHM of phonon as a function of pressure in Raman spectra.

Slight selenium substitution for tellurium in $Bi_2Te_3$ shifts the appearance of ETT from 3.2 to 2.5 GPa. In the meantime, the superconductivity presents follow the shifts of ETT at 3.2 GPa in $Bi_2Te_3$ and 2.4 GPa in $Bi_2Te_{2.1}Se_{0.9}$. Clearly, the superconductivity in this compound is pinned with the ETT, which indicates the ETT may play a crucial role in the appearance of superconductivity in $Bi_2Te_{3-x}Se_x$ at the initial phase of $R$-3m symmetry. These results underline that ETT highly promotes or even induces superconductivity in topological insulator $Bi_2Te_{2.1}Se_{0.9}$ in the low-pressure rhombohedral phase.

In summary, we have explored the high-pressure behaviors of $Bi_2Te_{2.1}Se_{0.9}$ using X-ray diffraction, Raman scattering, and electrical resistance measurements. By electrical resistance measurement, electrical resistance measurement can observe superconductivity at 2.4 GPa with a $T_c$ of about 6.6 K in $Bi_2Te_{2.1}Se_{0.9}$. X-ray diffraction and Raman experiments indicate that the first crystal phase transition occurs at about 8.4 GPa, much larger than the pressure value where superconductivity arises. Moreover, at about 2.5 GPa, the abnormal changes in the $c/a$ ratio and the FWHM of $A_{1g}^{1}$ mode indicate the occurrence of electronic topological transition (ETT). Compared with the well-investaged sample $Bi_2Te_3$, Our results suggest that ETT and superconductivity are highly coupled in $Bi_2Te_{2.1}Se_{0.9}$ at the low-pressure rhombohedral phase. Our results and finding clarifies the complex electronic structure and may shed light on the superconductivity mechanism in topological insulator $Bi_2Te_{2.1}Se_{0.9}$.

## Experimental Methods

The elements Bi, Te, and Se (99.5%, Alfa Aesar Co.) were weighted according to the molecular formula $Bi_2Te_{2.1}Se_{0.9}$. After being thoroughly grounded and pressed into a pellet, the mixture was loaded into a quartz tube sealed under a vacuum degree of $10^{-4}$ Pa. Then, the quartz tube was put into a muffle heated at the rate of 5 K/min to 1073 K and kept at the temperature for 12 hours to get the pure phase of $Bi_2Te_{2.1}Se_{0.9}$. The ingot was grounded to nanoscale powder and compacted by a vacuum hot-pressing furnace at a pressure of 55 MPa for 0.5 h. The sample was polycrystal with a relative density above 99.7% of the theoretical density. The crystal phase of the sample was confirmed by X-ray diffraction at room temperature using Cu Kalpha radiation.

The high-pressure electrical transport properties were investigated using the standard four-probe method in a diamond anvil cell (DAC) made of CuBe alloy. Slim Pt wires of 10 um in diameter were used as electrodes.[44] A T301 stainless steel gasket covered with fine cubic BN powder was used to insulate the electrode from the gasket. The low-temperature measurements were performed in Physical Property Measurement System (PPMS). High-pressure XRD and Raman measurements were performed using diamond anvil cells with a diameter culet of 300 um, and T301 stainless steel plates were used as gaskets. We used the standard ruby fluorescence method to calculate pressures in the sample chamber.[45] Renishaw inVia Raman microscope using standard backscattering geometry was used to carry out Raman measurements. A diode pumped solid state laser with the wavelength of 532 nm was applied to excite the sample, and the output power was 10 mW. The spectral resolution of the Raman system was around 1 $cm^{-1}$. In situ high-pressure angle-dispersive X-ray diffraction (ADXRD) experiments were performed on the 4W2 beamline at the High-Pressure Station of the Beijing Synchrotron Radiation Facility (BSRF) with a wavelength of 0.6199 Å. The average acquisition time was 300 s. The sample to detector distance and geometric parameters were calibrated using a $CeO_2$ standard. A MAR345 image plate detector was used to collect the diffraction patterns, and the two-dimensional XRD images were converted to one-dimensional intensity versus diffraction angle 2-theta patterns using FIT2D software.[46] High-pressure structural information was obtained using the Rietveld refinement method combined in the GSAS package.

## Acknowledgments

The authors want to show their great reverence and gratitude to Dr. Xiao-jia Chen (HPSTAR, Shanghai) and Dr. Viktor Struzhkin (Geophysical Laboratory, Carnegie Institution of Washington) for their helpful instructions and discussions. This work is supported by the financial supports from the National Natural Science Foundation of China (22074024), Key scientific research platforms and projects of Guangdong colleges and universities (2021ZDZX2043), and Natural Science Foundation of Guangdong Province (2214050009294). Medical Scientific Research Foundation of Guangdong

Province, China (No. A2021487), Science and Technology Program of Guangzhou, China (No. 202102020778). X-ray diffraction experiments were conducted at 4W2 beamline, Beijing Synchrotron Radiation Facility (BSRF) which is supported by Chinese Academy of Sciences. Portions of this work were performed at the 15U1 beamline at the Shanghai Synchrotron Radiation Facility (SSRF).

## Data availability

All data that support the findings of this study are available from the corresponding author upon reasonable request.

## Conflict of interest

The authors declare no conflict of interest.

## References

1. H. Zhang, C. X. Liu, X. L. Qi, X. Dai, Z. Fang, and S. C. Zhang, Topological insulators in $Bi_2Se_3$, $Bi_2Te_3$ and $Sb_2Te_3$ with a single Dirac cone on the surface. *Nature Phys.* **5**, 438 (2009).

2. X. L. Qi, and S. C. Zhang, Topological insulators and superconductors. *Rev. Mod. Phys.* **83**, 1057 (2011).

3. L. Miao, Y. Xu, W. Zhang, D. Older, S. A. Breitweiser, E. Kotta, H. He, T. Suzuki, J. D. Denlinger, R. R. Biswas, J. G. Checkelsky, W. Wu, and L. A. Wray, Observation of a topological insulator Dirac cone reshaped by non-magnetic impurity resonance. *npj Quant Mater* **3**, 29 (2018).

4. Y. S. Hor, A. J. Williams, J. G. Checkelsky, P. Roushan, J. Seo, Q. Xu, H. W. Zandbergen, A. Yazdani, N. P. Ong, and R. J. Cava, Superconductivity in $Cu_xBi_2Se_3$ and its implications for pairing in the undoped topological insulator. *Phys. Rev. Lett.* **104**, 057001 (2010).

5. Q. Li, D. E. Kharzeev, C. Zhang, Y. Huang, I. Pletikosić, A. V. Fedorov, R. D. Zhong, J. A. Schneeloch, G. D. Gu, and T. Valla, Chiral magnetic effect in $ZrTe_5$. *Nat. Phys.* **12**, 550 (2016).

6. Y. Xia, D. Qian, D. Hsieh, L. Wray, A. Pal, H. Lin, A. Bansil, D. Grauer, Y. S. Hor, R. J. Cava, and M. Z. Hasan, Observation of a large-gap topological-insulator class with a single Dirac cone on the surface. *Nature Phys*. **5**, 398 (2009).

7. Y. L. Chen, J. G. Analytis, J. H. Chu, Z. K. Liu, S. K. Mo, X. L. Qi, H. J. Zhang, D. H. Lu, X. Dai, Z. Fang, S. C. Zhang, I. R. Fisher, Z. Hussain, and Z.-X. Shen, Experimental realization of a three-dimensional topological insulator, $Bi_2Te_3$. *Science* **325**, 178 (2009).

8. D. Hsieh, Y. Xia, D. Qian, L. Wray, F. Meier, J. H. Dil, J. Osterwalder, L. Patthey, A. V. Fed
orov, H. Lin, A. Bansil, D. Grauer, Y. S. Hor, R. J. Cava, and M. Z. Hasan, Observation of ti
me-reversal-protected single-dirac-cone topological-insulator states in $Bi_2Te_3$ and $Sb_2Te_3$. *Phys. R
ev. Lett.* **103**, 146401 (2009).

9. S. Sasaki, M. Kriener, K. Segawa, K. Yada, Y. Tanaka, M. Sato, and Y. Ando, Topological sup
erconductivity in $Cu_xBi_2Se_3$. *Phys. Rev. Lett.* **107**, 217001 (2011).

10. L. Fu and E. Berg, Odd-parity topological superconductors: Theory and application to $Cu_xBi_2Se_3$.
*Phys. Rev. Lett.* **105**, 097001 (2010).

11. T. H. Hsieh and L. Fu, Majorana fermions and exotic surface Andreev bound states in topolog
ical superconductors: Application to $Cu_xBi_2Se_3$. *Phys. Rev. Lett.* **108**, 107005 (2012).

12. H. Peng, D. De, B. Lv, F. Wei, and C. W. Chu, Absence of zero-energy surface bound states
in $Cu_xBi_2Se_3$ studied via Andreev reflection spectroscopy. *Phys. Rev. B* **88**, 024515 (2013).

13. E. Lahoud, E. Maniv, M. S. Petrushevsky, M. Naamneh, A. Ribak, S. Wiedmann, L. Petaccia,
Z. Salman, K. B. Chashka, Y. Dagan, and A. Kanigel, Evolution of the Fermi surface of a do
ped topological insulator with carrier concentration. *Phys. Rev.B* **88**, 195107 (2013).

14. P. Das, Y. Suzuki, M. Tachiki, and K. Kadowaki, Spin-triplet vortex state in the topological su
perconductor $Cu_xBi_2Se_3$. *Phys. Rev. B* **83**, 220513(R) (2011).

15. N. Levy, T. Zhang, J. Ha, F. Sharifi, A. A. Talin, Y. Kuk, and J. A. Stroscio, Experimental ev
idence for s-wave pairing symmetry in superconducting $Cu_xBi_2Se_3$ single crystals using a scanni
ng tunneling microscope. *Phys. Rev. Lett.* **110**, 117001 (2013).

16. L. Fu and C. L. Kane, Superconducting proximity effect and Majorana fermions at the surface
of a topological insulator. *Phys. Rev. Lett.* **100**, 096407 (2008).

17. M. Kriener, K. Segawa, Z. Ren, S. Sasaki, and Y. Ando, Bulk superconducting phase with a f
ull energy gap in the doped topological insulator $Cu_xBi_2Se_3$. *Phys. Rev. Lett.* **106**, 127004 (201
1).

18. J. Zhu, J. L. Zhang, P. P. Kong, S. J. Zhang, X. H. Yu, J. L. Zhu, Q. Q. Liu, X. Li, R. C.
Yu, R. Ahuja, W. G. Yang, G. Y. Shen, H. K. Mao, H. M. Weng, X. Dai, Z. Fang, Y. S. Zha
o, and C. Q. Jin, Superconductivity in topological insulator $Sb_2Te_3$ induced by pressure. *Sci. R
ep.* **3**, 2016 (2013).

19. Y. Zhou, J. Wu, W. Ninga, N. Li, Y. Du, X. Chen, R. Zhang, Z Chi, X. Wang, X. Zhu, P. L
u, C. Jie, X. Wan, Z. Yang, J. Sun, W. Yang, M. Tian, Y. Zhang, and H. K. Mao, Pressure-in
duced superconductivity in a three-dimensional topological material $ZrTe_5$. *Proc. Natl. Acad. Sci.*
**113** 2904 (2016).

20. J. L. Zhang, S. J. Zhang, H. M. Weng, W. Zhang, L. X. Yang, Q. Q. Liu, S. M. Feng, X. C.
Wang, R. C. Yu, L. Z. Cao, L. Wang, W. G. Yang, H. Z. Liu, W. Y. Zhao, S. C. Zhang, X.
Dai, Z. Fang, and C. Q. Jin, Pressure-induced superconductivity in topological parent compou
nd $Bi_2Te_3$. *Proc. Natl. Acad. Sci.* **108**, 24 (2011).

21. J. Zhao, H. Liu, L. Ehm, Z. Chen, S. Sinogeikin, Y. Zhao, and G. Gu, Pressure-induced disor
dered substitution alloy in $Sb_2Te_3$. *Inorg. Chem.* **50**, 11291 (2011).

22. Y. Li and Z. Xu, Exploring topological superconductivity in topological materials. *Adv. Quantu
m Technol.* **2**, 1800112 (2019).

23. S. K. Mishra, S. Satpathy, and O. Jepsen, Electronic structure and thermoelectric properties of
bismuth telluride and bismuth selenide. *J. Phys.: Condens. Matter.* **9**, 461 (1997).

24. I. M. Lifshitz, Anomalies of electron characteristics of a metal in the high pressure region. *Sov.
Phys. JETP (USSR)* **11**, 1130 (1960).

25. A. Polian, M. Gauthier, S. M. Souza, D. M. Trichês, J. C. Lima, and T. A. Grandi, Two-dimensional pressure-induced electronic topological transition in $Bi_2Te_3$. *Phys. Rev. B* **83**, 113106 (2011).

26. N. Sakai, T. Kajiwara, K. Takemura, S. Minomura, and Y. Fujii, Pressure-induced phase transition in $Sb_2Te_3$. *Solid State Commun.* **40**, 1045 (1981).

27. M. K. Jacobsen, R. S. Kumar, A. L. Cornelius, S. V. Sinogeiken, and M. F. Nicol, High pressure X-ray diffraction studies of $Bi_{2-x}Sb_xTe_3$ (x = 0,1,2). *AIP Conf. Proc.* **955**, 171 (2007).

28. C. Zhang, L. Sun, Z. Chen, X. Zhou, Q. Wu, W. Yi, J. Guo, X. Dong, and Z. Zhao, Phase diagram of a pressure-induced superconducting state and its relation to the Hall coefficient of $Bi_2Te_3$ single crystals. *Phys. Rev. B* **83**, 140504(R) (2011).

29. C. Chen, S. He, H. Weng, W. Zhang, L. Zhao, H. Liu, X. Jia, D. Mou, S. Liu, J. He, Y. Peng, Y. Feng, Z. Xie, G. Liu, X. Dong, J. Zhang, X. Wang, Q. Peng, Z. Wang, S. Zhang, F. Yang, C. Chen, Z. Xu, X. Dai, Z. Fang, and X. J. Zhou, Robustness of topological order and formation of quantum well states in topological insulators exposed to ambient environment. *Proc. Natl. Acad. Sci.* **109**, 3694 (2012).

30. K. Matsubayashi, T. Terai, J. S. Zhou, and Y. Uwatoko, Superconductivity in the topological insulator $Bi_2Te_3$ under hydrostatic pressure. *Phys. Rev. B* **90**, 125126 (2014).

31. C. Buzea, T. Tachiki, K. Nakajima, and T. Yamashita, The origin of resistance peak effect in high-temperature superconductors-apparent $T_c$ anisotropy due to $J_c$ anisotropy. *IEEE Trans. Appl. Supercond.* **11**, 3655 (2001).

32. H. Yamamoto, M. Ikeda, and M. Tanaka, Giant resistivity anomaly in A15 Nb3(Ge, Si) superconductive films with compositionally modulated superstructure. *Jpn. J. Appl. Phys*. **24,** 314 (1985).

33. J. Zhang, Z. Peng, A. Soni, Y. Zhao, Y. Xiong, B. Peng, J. Wang, M. S. Dresselhaus, and Q. Xiong, Raman spectroscopy of few-quintuple layer topological insulator $Bi_2Se_3$ Nanoplatelets. *Nano Lett.* **11**, 2407 (2011).

34. J. R. Wiese and L. Muldawer, Lattice constants of $Bi_2Te_3$-$Bi_2Se_3$ solid solution alloys. *J. Phys. Chem. Solids* **15**, 13 (1960).

35. J. R. Drabble and L. J. Goodman, Chemical bonding in bismuth telluride. *Phys. Chem. Solids* **5**, 142 (1958).

36. A. Soni, Z. Yanyuan, Y. Ligen, M. K. K. Aik, M. S. Dresselhaus, and Q. Xiong, Enhanced thermoelectric properties of solution grown $Bi_2Te_{3-x}Se_x$ nanoplatelet composites. *Nano Lett.* **12**, 1203 (2012).

37. H. Köhler, and C. R. Becker, Optically active lattice vibrations in $Bi_2Se_3$. *Phys. Status Solidi B* **61**, 533 (1974).

38. A. Torabi, Y. Song, and V. N. Staroverov, Pressure-induced polymorphic transitions in crystalline diborane deduced by comparison of simulated and experimental vibrational spectra. *J. Phys. Chem. C* **17**, 2210 (2013).

39. R. Vilaplana, O. Gomis, F. J. Manjón, A. Segura, E. Pérez-González, P. Rodríguez-Hernández, A. Muñoz, J. González, V. Marín-Borrás, V. Muñoz-Sanjosé, C. Drasar, and V. Kucek, High-pressure vibrational and optical study of $Bi_2Te_3$. *Phys. Rev. B* **84**, 104112 (2011).

40. A. Bera, K. Pal, D. V. S. Muthu, S. Sen, P. Guptasarma, U. V. Waghmare, and A. K. Sood, Sharp Raman anomalies and broken adiabaticity at a pressure induced transition from band to topological insulator in $Sb_2Se_3$. *Phys. Rev. Lett.* **110**, 107401 (2013).

41. P. P. Kong, J. L. Zhang, S. J. Zhang, J. Zhu, Q. Q. Liu, R. C. Yu, Z. Fang, C. Q. Jin, W. G. Yang, X. H. Yu, J. L. Zhu, and Y. S. Zhao, Superconductivity of the topological insulator $Bi_2$ $Se_3$ at high pressure. *J. Phys.: Condens. Matter.* **25**, 362204 (2013).

42. K. Kirshenbaum, P. S. Syers, A. P. Hope, N. P. Butch, J. R. Jeffries, S. T. Weir, J. J. Hamlin, M. B. Maple, Y. K. Vohra, and J. Paglione, Pressure-induced unconventional superconducting phase in the topological insulator $Bi_2Se_3$. *Phys. Rev. Lett.* **111**, 087001 (2013).

43. V. V. Struzhkin, Y. A. Timofeev, R. J. Hemley, and H. K. Mao, Superconducting $T_c$ and electron-phonon coupling in Nb to 132 GPa: Magnetic susceptibility at megabar pressures. *Phys. Rev. Lett.* **79**, 4262 (1997).

44. A. G. Gavriliuk, A. A. Mironovich, and V. V. Struzhkin, Miniature diamond anvil cell for broad range of high pressure measurements. *Rev. Sci. Instrum*. **80**, 043906 (2009).

45. H. K. Mao, J. A. Xu, and P. J. Bell, Calibration of the ruby pressure gauge to 800 kbar under quasi‐hydrostatic conditions. *Geophys. Res* **91**, 4673 (1986).

46. A. P. Hammersley, S. O. Svensson, M. Hanfland, A. N. Fitch, and D. Hausermann, Two-dimensional detector software: from real detector to idealised image or two-theta scan. *High Pres. Res.* **14**, 235 (1996).

## Figure Captions

**Figure. 1.** Temperature-dependent resistance of $Bi_2Te_{2.1}Se_{0.9}$ at various pressures in the temperature range 2-14 K. The inset shows the resistance of $Bi_2Te_{2.1}Se_{0.9}$ at 1 atm at the temperature of 2-300 K.

**Figure 2.** (a) The Raman spectra of $Bi_2Te_{2.1}Se_{0.9}$ at different pressures. (b) Pressure dependence of the Raman modes. (c) The full width at half maximum for mode versus pressure.

**Figure 3.** Rietveld refinements of the XRD patterns at different pressures, (a) 0.6, (b) 13.3, (c) 21.2, and (d) 36.3 GPa, respectively. The corresponding crystal structure of the four phases is shown in the insets. The solid red lines and open black circles represent the Rietveld fits for the simulated and observed data, respectively The solid black lines at the bottom are the residual intensities. The vertical bars indicate the peak positions. XRD pattern in panel (c) is a mixture of phases II, III, and IV.

**Figure. 4.** The pressure-dependent (a) lattice parameters a and c axes and (b) $c/a$ values in phase I. (c) The experimental volume per atom versus pressure for phases I, II, III, and IV. The reduced pressure (d) H, (e) $H_a$, (f) $H_c$ versus Eulerian strain $f_E$. Linear fits are performed to the data points in two

regions for (d), (e), and the whole region for (f). The c parameter is unaffected by the ETT.

**Figure. 5.** The evolutions of Tconset and the crystal structure as a function of pressure. The empty circle at about 1.0 GPa is the depressing data.

## Figures

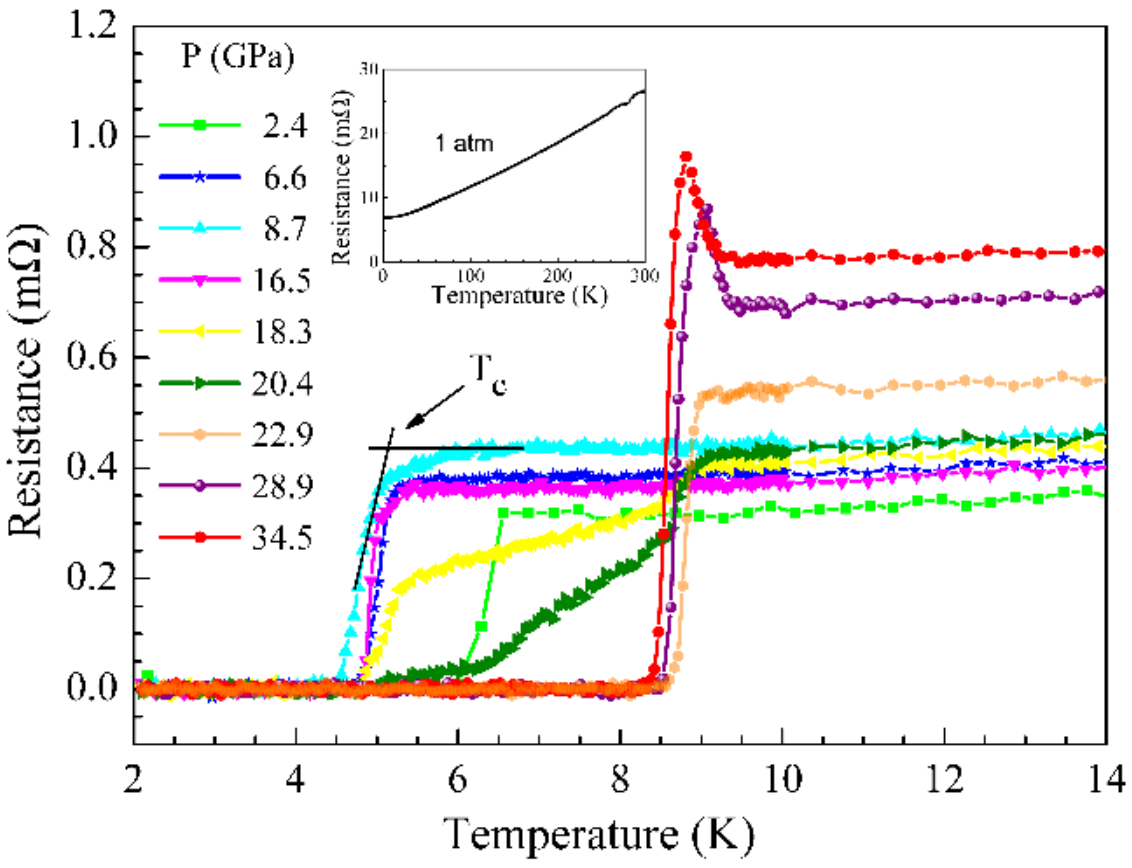

Figure 1

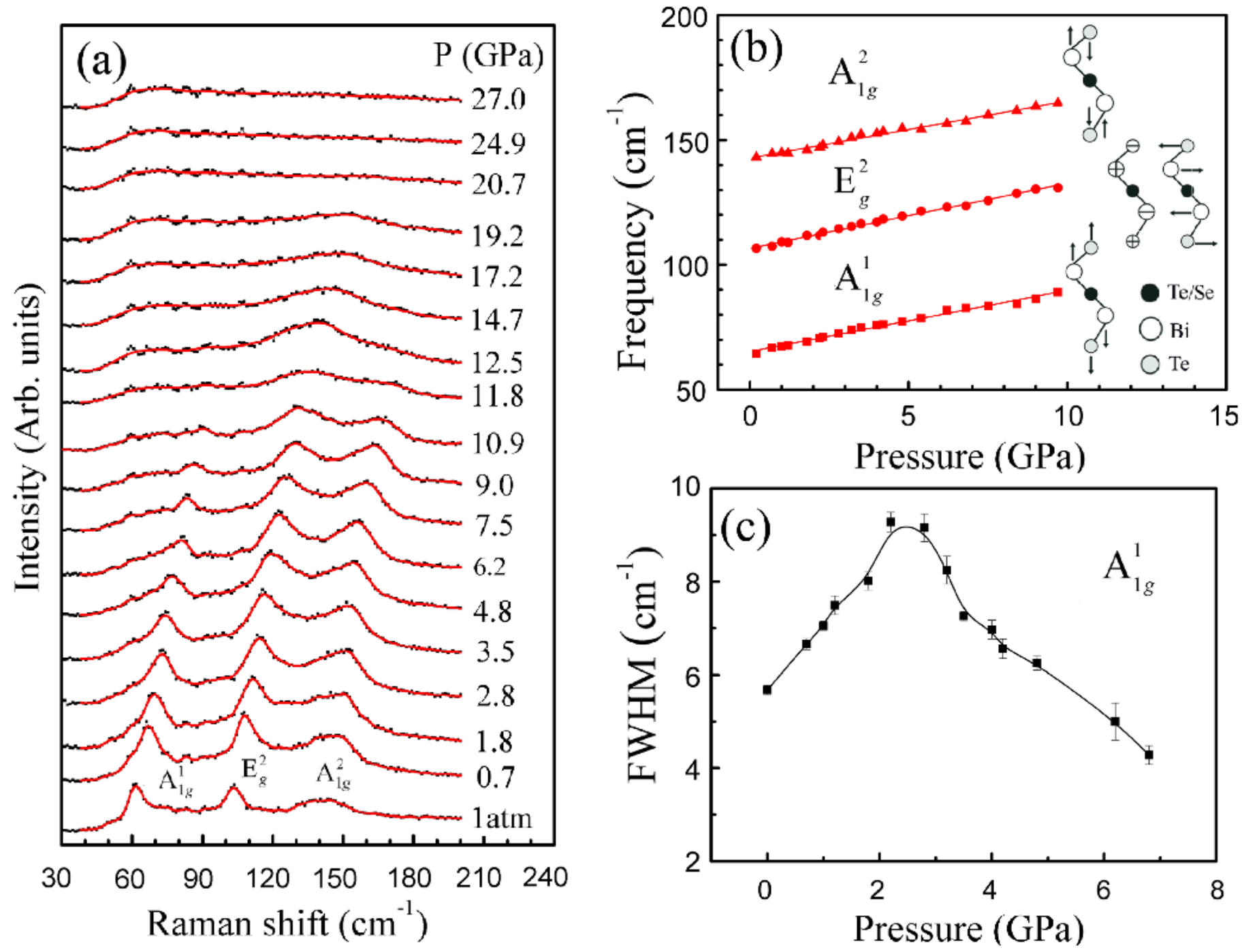

Figure 2

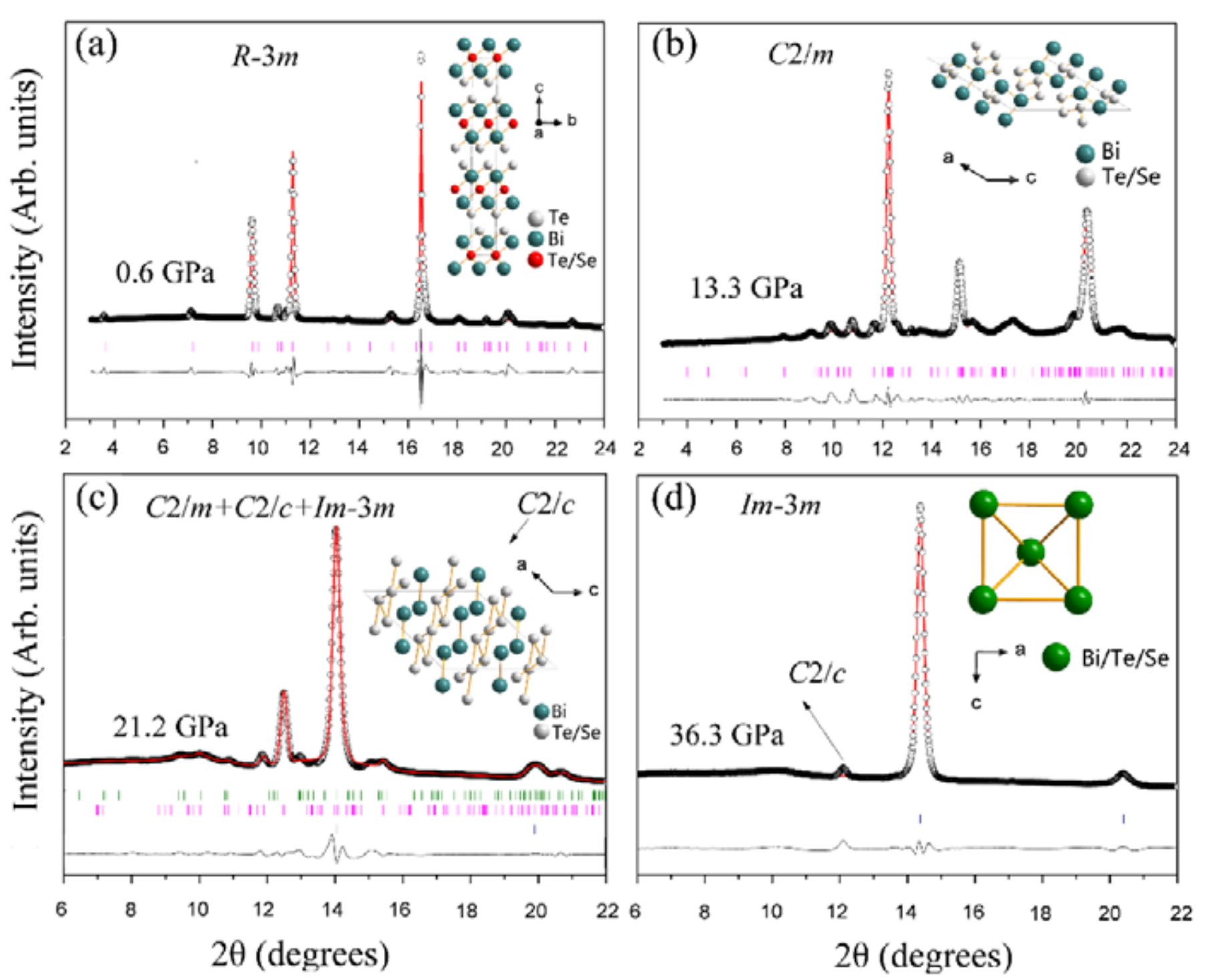

Figure 3

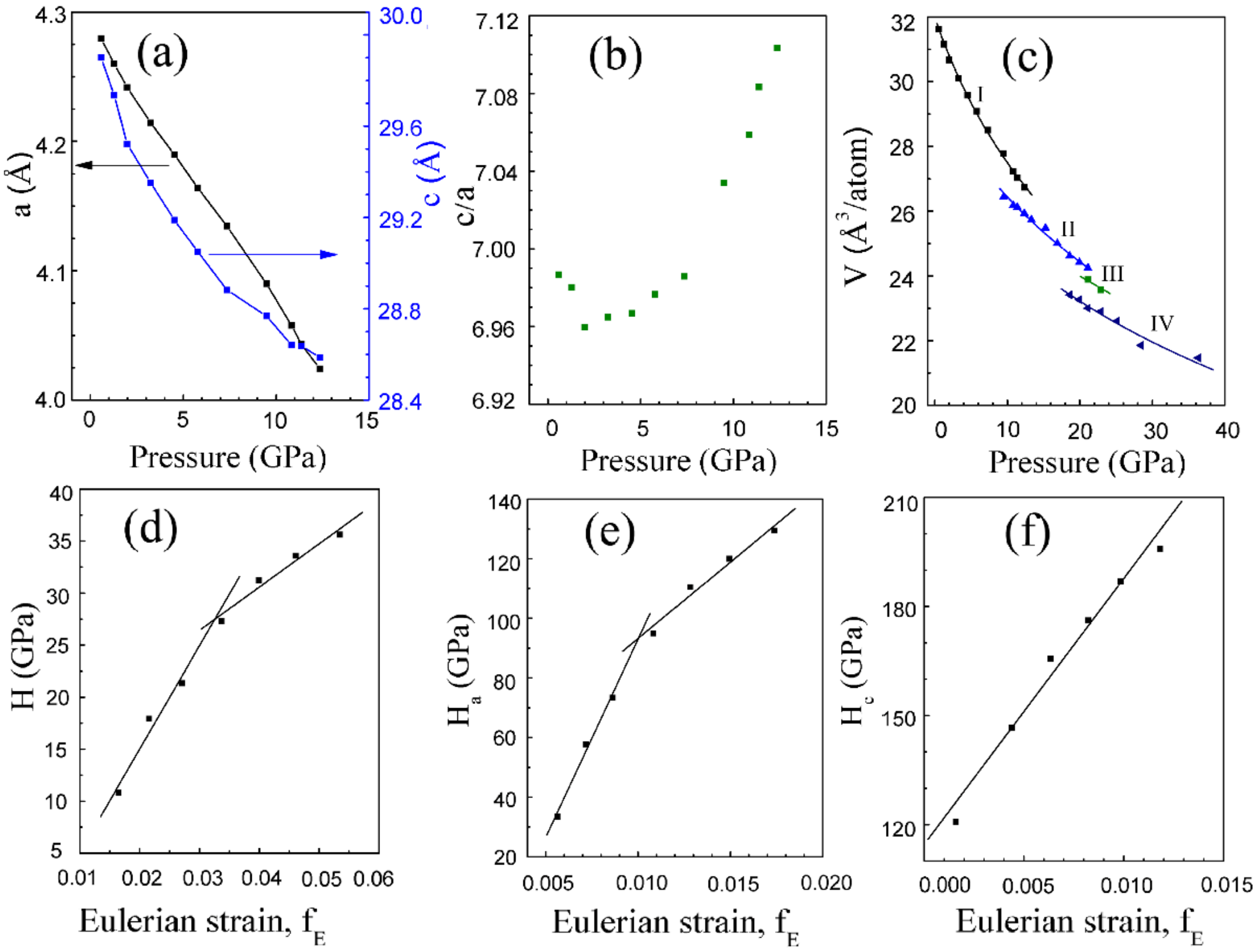

Figure 4

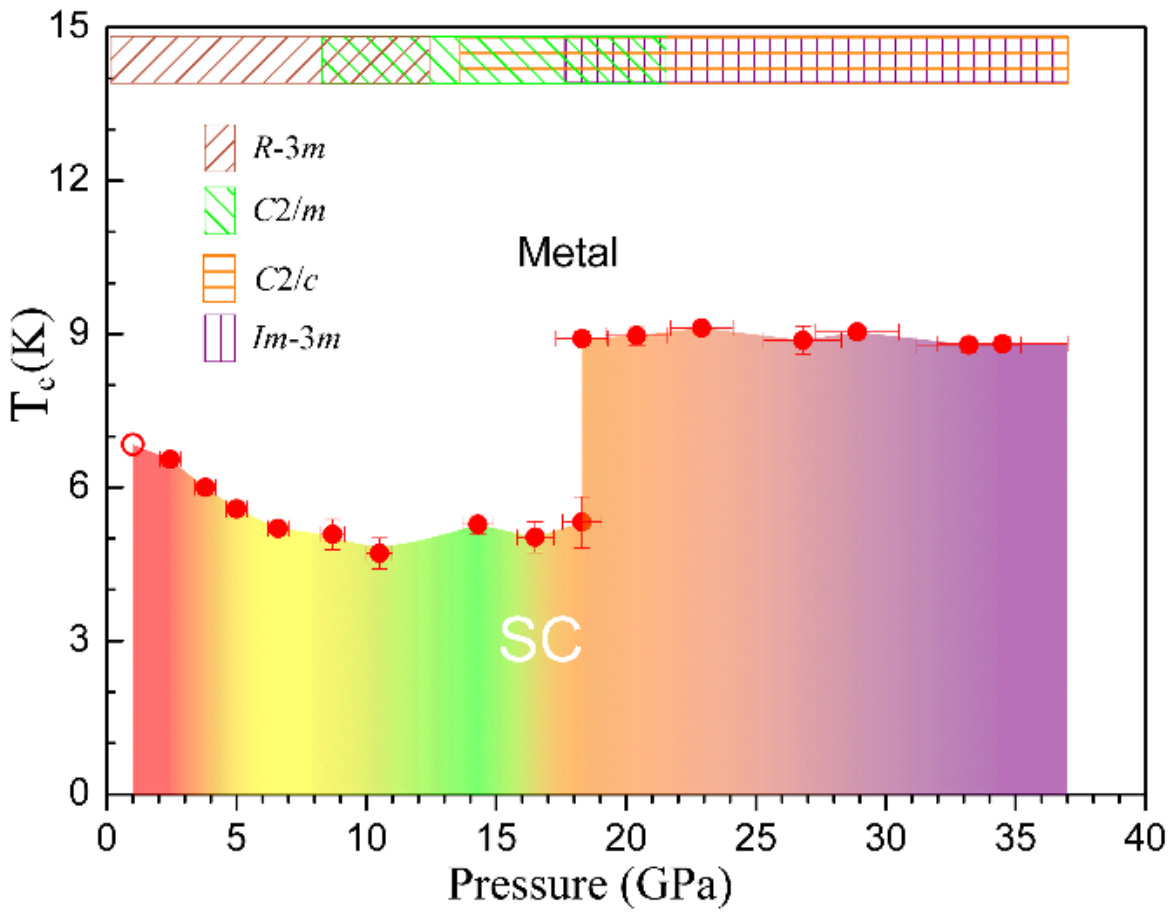

Figure 5

## Supporting Information

### Figure Captions

**Figure. S1.** Representative X-ray diffraction patterns of $Bi_2Te_{2.1}Se_{0.9}$ at different pressures. The peaks marked by asterisks, rhombuses, and solid circles indicate the start of new phases. The up-facing and down-facing arrows indicate that the intensities of the peaks become stronger or weaker with increasing pressure.

### Figures

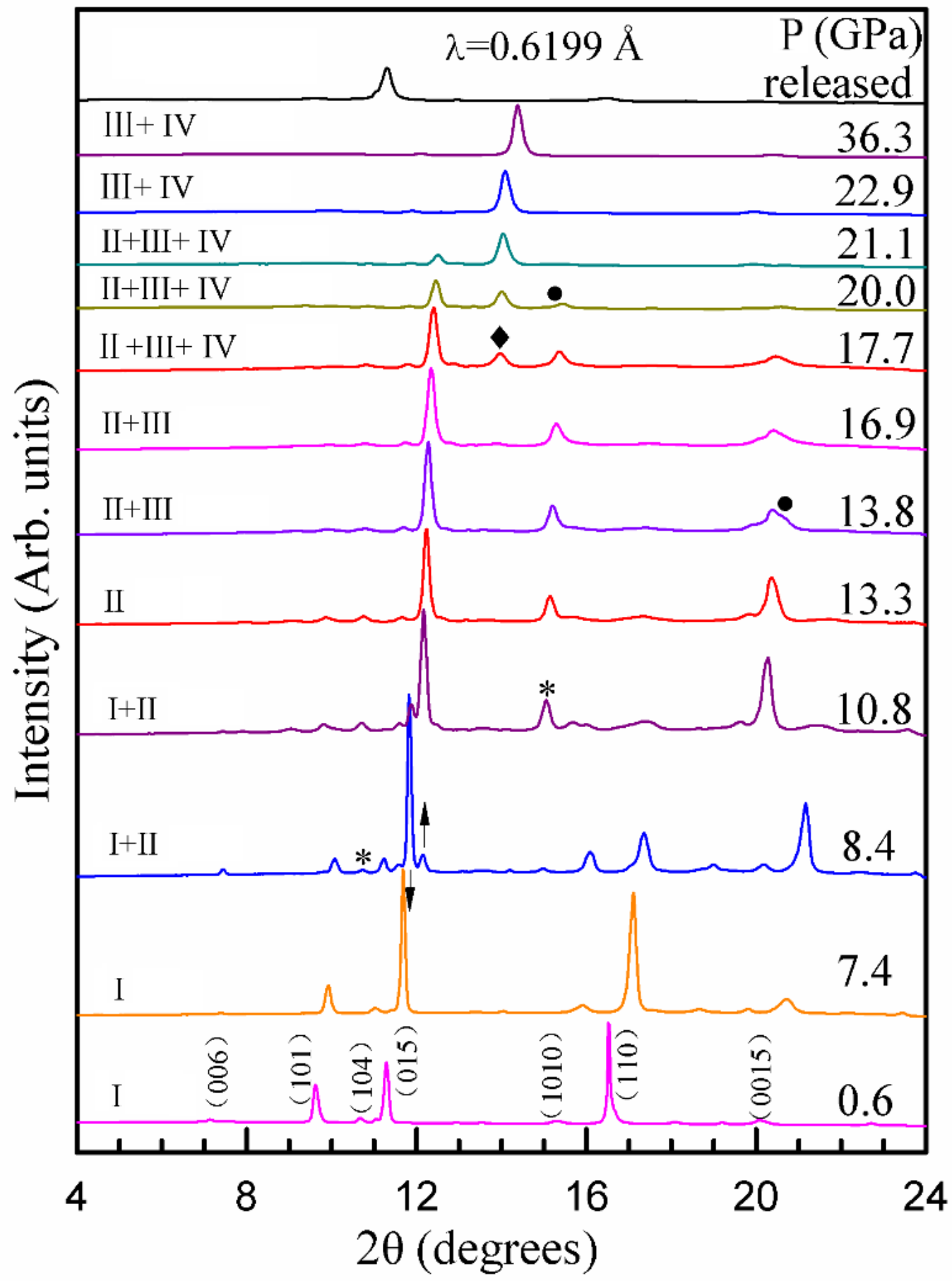

Figure S1

**For Table of Contents Use Only**

# Pressure-induced electronic topological transition and superconductivity in topological insulator $Bi_2Te_{2.1}Se_{0.9}$

*Lei Kang,[1] Zi-Yu. Cao[2*], and Bo Wang[3]*

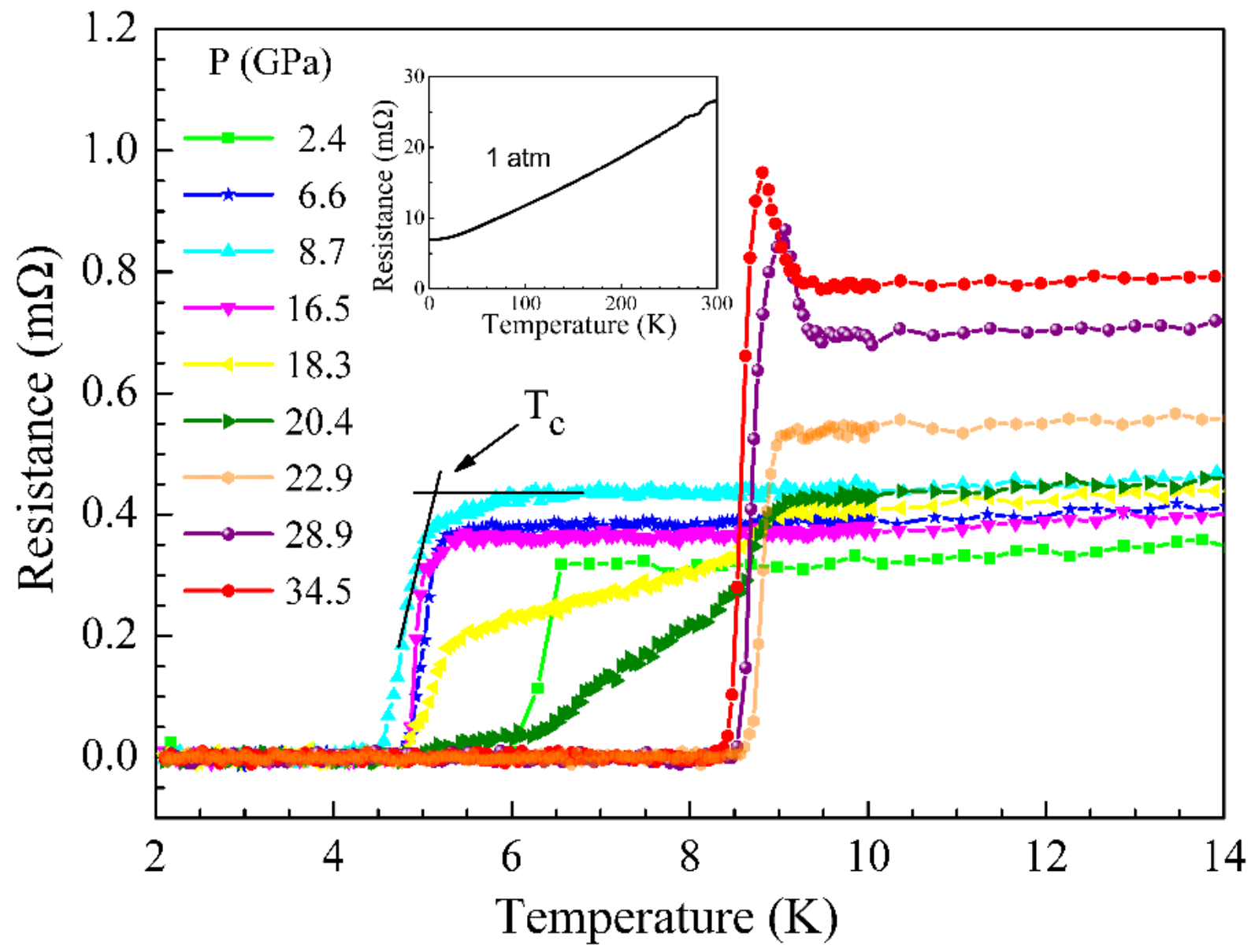